\documentclass[11pt, showpacs,preprintnumbers,amsmath,amssymb]{revtex4}

\usepackage{graphicx}
\usepackage{bm}
\begin{document}
\title{Complete synchronization in coupled Type-I neurons}
\author{Nishant Malik$^1$}
\author{B. Ashok$^2$}
\author{J. Balakrishnan$^3$}
\altaffiliation{Corresponding author}
\email{janaki05@gmail.com, jbsp@uohyd.ernet.in }
\affiliation{
$^1$Potsdam Institute for Climate Impact Research, Telegrafenberg, 14412 Potsdam, Germany.}
\affiliation{$^2$Advanced Centre for Research in High Energy Materials (ACRHEM),\\
University of Hyderabad, Central University PO, Gachi Bowli, Hyderabad - 500 046, India.}
\affiliation{ $^3$School of Physics, University of Hyderabad, Central University PO,
Gachi Bowli, Hyderabad - 500 046, India.}
\begin{center}
{\underline {{\bf Published in:} ~ {\large {\em Pramana - Journal of Physics} {\bf 74}, 
189 (2010)}}}
\end{center}
\begin{abstract}
For a system of type-I neurons bidirectionally coupled through a nonlinear
feedback mechanism, we discuss the issue of noise-induced complete
synchronization (CS). For the inputs to the neurons, we point out that the rate
of change of instantaneous frequency with the instantaneous phase of the
stochastic inputs to each neuron matches exactly with that for the other in the
event of CS of their outputs. Our observation can be exploited in practical
situations to produce completely synchronized outputs in artificial devices.
For excitatory-excitatory synaptic coupling, a functional dependence for the
synchronization error on coupling and noise strengths is obtained. Finally we
report an observation of noise-induced CS between non-identical neurons coupled
bidirectionally through random non-zero couplings in an all-to-all way in a
large neuronal ensemble.
\end{abstract}

\maketitle
\vspace*{0.5cm}

\section{Introduction}
Synchronous phenomena abound in nature and in our daily lives and have been
studied from centuries past, right from Huygen's observations of synchronizing
clocks. Various kinds of synchronous phenomena occur and have been identified 
(see for example~\cite{pecora,hanselsompolinsky,cerdeira,vreeswijk,hansel,
ermentrout,borgerskopell,borgerskopell2,izhikevich,cerdeira2,cerdeira3,
sinha,jampa})
(other references on the subject may also be found in ~\cite{pikovsky}). 
Among these complete synchronization (CS) is one of the
most interesting since the phase, frequency and amplitude of a subsystem
all coincide with those of the other subsystem it is coupled to.
It is seen therefore that in CS the trajectories of the coupled elements
match exactly. CS is known to occur in identical systems, and was first
demonstrated in chaotic systems in \cite{pecora}.
Neurons and neuronal networks have been a subject of frequent theoretical and
experimental studies~\cite{tuckwell}. Synchronization of neural activity has  
elicited a great deal of interest since it is believed that such phenomena enable 
cognitive tasks such as feature extraction and recognition to be performed. 
Hodgkin and Huxley classified neuron excitability mechanisms broadly into two 
classes~\cite{hodgkin}: in type-II neurons the transition from a quiescent state 
to a periodically spiking state occurs through a Hopf bifurcation with a finite 
nonzero oscillation frequency. 
In type-I neurons, oscillations emerge through a saddle-node bifurcation on an 
invariant circle. As the bifurcation parameter changes, the stable and the 
unstable fixed points coalesce and then disappear, leaving a large amplitude 
stable periodic orbit. This is a global bifurcation and the frequency of the 
global loop can be arbitrarily small. 
Since axonal excitability patterns of mammalian neurons fall under the type-I 
class, it is but natural that this class has received special attention in the 
literature. Various observations have been made on type-I neurons, some prominent 
points of which are as follows. Equations for type-I neuronal dynamics can be 
reduced to the canonical normal form for a saddle-node bifurcation~\cite{rinzel}.
Repetitive firing occurs in the parameter regime when the system is in the
close proximity of a saddle-node bifurcation on an invariant circle. Hansel
{\em et al}~\cite{hansel} and Ermentrout have shown~\cite{ermentrout} that such
neurons coupled via a certain class of time-dependent synaptic conductances are
difficult to synchronize. B{\"o}rgers and Kopell made further 
investigations of such coupled systems. In particular they discussed the effects 
of random connectivity on synchronization and the PING mechanism in networks of
excitatory (E) and inhibitory (I) neurons both in the presence and in the
absence of external noise~\cite{borgerskopell,borgerskopell2}.

In this work, we present some computer studies of generic type-I neurons
coupled via synaptic conductances such as those considered
in~\cite{borgerskopell,borgerskopell2,izhikevich}, which are governed by
ordinary differential equations and which depend upon the outputs of the
presynaptic neurons, and are subject to weak additive Gaussian white noise. We
consider both excitatory-excitatory (EE) and inhibitory-excitatory (IE)
bidirectional couplings and show that in certain regimes of the coupling
constants and inputs, the system of coupled neurons shows complete
synchronization (CS). The issue of CS in type-I neurons was not discussed in
~\cite{borgerskopell,borgerskopell2}. As discussed in Section-2, largest
Lyapunov exponents are shown to not adequately give information about CS in the
system. We make an observation on the inputs to the neurons (and which are
also modulated by the feedback in the system): we point out that the variation
of the instantaneous frequency of the input received at each neuron with the
instantaneous phase of the input it receives exactly coincides with that of the
other neuron, in the event of complete synchronization of their outputs. It
will be noted that in the presence of noise and feedback, this is not a trivial
statement. We discuss the utility of this result in practical situations.\\
In general, for EE synapses, our results indicate that when a common,
externally applied constant input is used to perturb two bidirectionally
coupled type-I neurons having identical coupling strength magnitudes and
synaptic rise \& decay times, weak noise induces them to exhibit CS upto a
critical value of the coupling strength $g_c$. For coupling strengths larger
than $g_c$, we find the system de-synchronizes through a power-law before
locking on to a partially synchronized state for larger coupling strengths. We
obtain a functional dependence for the synchronization error for neuronal
outputs on coupling and noise strengths, in the regimes leading to partial
synchronization. Such functional dependencies have not been reported in the
literature yet, to the best of our knowledge. In the noiseless case for
identical EE neurons separated by different initial conditions, we observe that
the antiphase states are stable in agreement with~\cite{ermentrout} and become
completely in-phase in the presence of noise. For just two neurons with IE
coupling, noise does not induce complete synchrony. In an ensemble of 200
non-identical neurons however, we find unexpectedly that noise-induced CS is
possible with all-to-all bidirectional IE random couplings.
\section{Coupled type-I neurons in the presence of Gaussian white noise}
In a system of $n$ neurons, the activity of the $i$th neuron is described by a
variable $x_i$ which can be related to the membrane conductance. Its dynamics
is represented by
\begin{equation}
\dot x_i = qx_i^2 + I_i , ~~~~I_i = \beta_i +
\sum_{j=1}^{n}\alpha_jg_{ji}s_{ji}~,
\end{equation}
where $q$ denotes an inverse time constant for the membrane potential, $I_i$ denotes
its total input comprising of a constant external input $\beta_i$
and the contributions from the presynaptic neurons, with $i=1,\dots n$.
$s_{ij}$ is the synaptic gating variable and represents the fraction of
ion channels open in the $j$th presynaptic neuron. $g_{ij}$ is the measure of the
strength of the synapse from neuron $i$ to neuron $j$; we have taken $g_{ii}=0$.
When $I_j >0$ this equation has no fixed points. Any initial condition tends
to infinity in a finite time. To avoid this blow-up of solutions, a nonlinear
transformation to new variables $\theta_i$ may be made~\cite{ermentrout}:
~$x_i = \tan{\theta_i/2}$ which maps the real line onto a circle.
Eqn.(2) then becomes
\begin{equation}
\frac{d\theta_i}{dt} = q(1 - \cos \theta_i) +
(\beta_i + \sum_{j=1}^{n}\alpha_jg_{ji}s_{ji}(\theta_j))(1 + \cos \theta_i)
\end{equation}
The point ~$x= \infty$ then gets mapped to the point $\theta = \pi$ and is
interpreted as firing of a spike.
We set $q^{-1}=1$ and work in the parameter regime in which $I/q<<1$ so that
the width of the spikes turn out to be in milliseconds as in real
neurons~\cite{borgerskopell,borgerskopell2} in these units.
The $s_{ij}(t)$ evolve in time according to the differential equation which
was considered in~\cite{borgerskopell,borgerskopell2,izhikevich}
\begin{equation}
\frac{ds_{ij}}{ dt}= -\frac{s_{ij}}{\tau_{ij}} + e^{-\eta(1+\cos \theta_i)}
\frac{1-s_{ij}}{\tau_R}
\end{equation}
where $\tau_{ij}$ denotes the synaptic decay time and $\tau_R$ the synaptic rise time.
The values of $s_{ij}$ always
lie in the range 0 to 1, reaching the maximal value when the neuron spikes.
The synapse is an excitatory one if $\alpha_j=+1$, and $-1$ if it models an
inhibitory synapse. Lest there be any confusion, we would like to clarify
at the outset that when we refer to {\it identical} neurons, we mean neurons
that have the same nature of synaptic coupling (EE) and have same coupling strengths
$g_{12} = g_{21}$, have the same value of $\tau_R$, $\tau_{12} = \tau_{21}$, and receive the
same constant input $\beta_1 = \beta_2$, while their initial conditions differ by very little.\\
Transmembrane voltage and neuronal firing can be affected by various sources of
neuronal noise, but predominantly by synaptic noise ~\cite{koch}. The synaptic
noise itself occurs due to several factors, but chief among them is the
synaptic bombardment at the inputs through the large number of neuronal
connections, with each input spike adding a random contribution. We model this
through an additive Gaussian white noise added to the neuronal input $I_i$. We
study the dynamics of a system of two such neurons coupled bidirectionally as
depicted in Fig.(1) and subject to Gaussian white noise $\xi(t)$ with the
following properties: $\langle \xi(t)\rangle = 0$ ~, ~$\langle
\xi(t)\xi(t')\rangle = 2 \sigma \delta(t-t')$, where the stochastic variables
are taken to obey Stratonovich calculus. Addition of Gaussian white noise $\xi$
to eqn.(1), manifests as multiplicative noise in eqn.(2) because of the change
of variables to $\theta$, so that the equations now take the form
\begin{equation}
\frac{d\theta_i}{dt} = (1 - \cos \theta_i) +
(\beta_i + \sum_{j=1}^{n}\alpha_jg_{ji}s_{ji}(\theta_j)+ \xi(t) )(1 + \cos \theta_i)
\end{equation}
Eqns.(3) for the $s_{ij}(\theta_i)$ define the feedback regulating the
activity of the $j$th neuron since $s_{ij}$  depends upon $\theta_i$ which in turn
depends upon $\theta_j$ ~($i \ne j$) ~via $s_{ji}(\theta_j)$.
The feedback increments or decrements the constant external input $\beta_i$ received
by neuron $i$. Thus the control parameter $I_i$ acquires time dependence through
the dynamical variables.

As in~\cite{lim}, we consider the neuronal output to be described by the variable
$u_i=(1-\cos \theta_i)/2$
as its time evolution pattern resembles that of a membrane potential in
real neurons. This transformation maps the resting point $x_i = 0$ corresponding
to $\theta_i =0$ to $u_i=0$, and the spiking point $\theta_i= \pi $ to $u_i=1$
via the relation $u_i = x_i^2/(1 + x_i^2)$.
We choose to work with these variables as we get some new and interesting insights
upon the dynamics underlying the phenomenon of complete synchronization.
In terms of these variables, eqns.(4) and (3) become:
\begin{eqnarray}
\dot{u_i}&=&2(u_i+(\beta_i+\sum_{j=1}^{N}\alpha_jg_{ji}s_{ji} + \xi)(1-u_i))\sqrt{u_i(1-u_i)}\nonumber\\
\dot{s}_{ij}&=&-\frac{s_{ij}}{\tau_{ij}}+\exp{(-2\eta(1-u_j))}\frac{(1-s_{ij})}{\tau_R}
\end{eqnarray}
Numerous studies in the literature have reported the phenomenon of complete
synchronization in various coupled systems~\cite{pecora,hanselsompolinsky,
cerdeira,vreeswijk,hansel,ermentrout,borgerskopell,borgerskopell2,izhikevich,
cerdeira2,cerdeira3,sinha,jampa,pikovsky,anishchenko,zhou,zhou2,wang,toral,he}. 
An adequately satisfying explanation of why and under what conditions CS can 
occur for systems with more complicated couplings, such as, for instance, that
described by eqn.(5), is however, still lacking in our opinion.\\
In particular, in ~\cite{zhou,zhou2} the authors study noise induced CS in systems
subjected to a common additive white noise and show that a necessary condition
for CS is the existence of a significant contraction region in phase space. The
systems studied in ~\cite{zhou,zhou2} were Lorenz and Rossler systems which are far
more amenable to analytical treatment than the equations above in eqn.(5). \\
Since excitability in type-I neurons results from a saddle node on an invariant
circle bifurcation, complete synchronization of two uncoupled neurons by common
noise alone can be expected because of the existence of a contraction region
close to the stable manifold of the saddle. On the other hand, when two such
neurons are coupled together as in eqn.(5), the existence and nature of a
contraction region would depend upon the eigenvalues of the Jacobian at the
fixed points of the coupled system. However for eqns.(5), an analytical study
becomes difficult since the Jacobian becomes singular at the fixed points. We
therefore perform some computer studies on the system to learn more about the
underlying dynamics. Moreover, since as we show below, Lyapunov exponents need
not adequately give information about CS, we seek other explanations for
occurrence of CS.\\
As in ~\cite{zhou,zhou2}, we define CS between the activities of neurons 
1 and 2 as a vanishing value for the quantity $\langle |u_1-u_2|\rangle$ which 
is the synchronization error averaged over all iterations.

Largest Lyapunov exponents for the system in eqn.(5) in the presence of noise
for both EE and IE couplings were calculated following
~\cite{parkerchua,eckmann,gao} and are shown in fig.(2). To incorporate noise
in the numerical calculations, the stochastic Runge-Kutta-4
method~\cite{rumelin,hansen,kloeden} was used. We note that in both EE and IE
cases, the largest Lyapunov exponent $\lambda_1$ becomes more negative on the
addition of noise. We observe also that for some intermediate values of the
coupling constants (such as $g_{12}=g_{21}=0.5$), the $\lambda_1$ values could
be larger, i.e., less negative, than those for lower coupling strengths (e.g.,
$g_{12}=g_{21}=0.3$). For EE coupling, $\lambda_1$ is almost always less than
or equal to zero. In the case of IE coupling however, we find that for smaller
noise-strengths and smaller couplings, for small $\beta$, $\lambda_1$
fluctuates between positive and negative values in the presence of noise. This
happens because of the oscillation of the bifurcation parameter (total input)
between two regimes, depending upon the relative strengths of $\beta_i$ and
$\alpha_jg_{ji}s_{ji}$, since $\alpha_j=-1$ for neuron $j$. Hence calculation
of Lyapunov exponents may not adequately give information about CS or large
windows of zero synchronization error, such as, for example, for the situation
shown in Fig.(3), though they may certainly show the emergence of a definite
order in the presence of noise and possible synchronization between the coupled
units.

Indeed, CS is expected to occur between identical systems and finding CS
between non-identical oscillators would be unusual. As we describe later,
however, we do find noise-induced CS in a system of 200 non-identical
oscillators with random all-to-all couplings. Equally intriguingly, we find
that there are parameter regimes where identical oscillators do not show CS at
all, though the Lyapunov exponents shown in Fig.(2a) remain negative. Negative
transverse Lyapunov exponents are widely accepted as characterizing CS, but
their calculation for the system of 200 neurons, with a nontrivial feedback
mechanism for phase resetting, as in eqn.(5) is a difficult task and we have
not attempted it here. In our case, even for $n=2$, if we were to define new
variables: $z = \frac{(u_1-u_2)}{2}$ and $w = \frac{(u_1+u_2)}{2}$, expressing
the synchronization error dynamics through $\dot{z}$ in terms of $z$ and $w$
alone is not at all
 straightforward for the equations in (5).\\
Moreover when each of the subsystems has a saddle-node-on-an-invariant circle
bifurcation in the uncoupled limit, one could expect windows of intermittent
firing patterns. In the coupled system, this is indeed observed in some
parameter regimes, interspersed with large windows showing zero synchronization
error, even in identical oscillators (EE case) for lower noise strengths
(Fig.(3)). Hence we believe that negative largest Lyapunov exponents alone may
not constitute conclusive proof for predicting noise-induced CS in coupled
neurons.\\
\indent We therefore looked for other indicators which could help in
understanding the mechanism of CS better in systems with feedback, such as in
eqns.(5). We found one such simple indicator for CS in the context of the model
under study and which we will now describe. The same methods and analysis
should also hold for getting information on CS in any other system. Since
synchronous activity is brought about by a common input or through mutual
interactions and since these include components which are highly random, we
study the {\em instantaneous} values of the sum total {\em of the inputs}
received by each unit of a coupled system. We first set up a framework for this
purpose and then provide a physical motivation and explanation for
understanding CS through this indicator.
\section{Instantaneous phase - Instantaneous frequency variations of the Neuronal Inputs \& CS}
We construct the analytical signal~\cite{pikovsky,anishchenko}:
$\Gamma_i(t)=I_i(t)+iH(I_i(t))= B_i(t)e^{i\rho_i(t)}$~ for the inputs $I_i$ and
similarly $w(t)=u(t)+iH(u(t))= R(t)e^{i\phi(t)}$ for $u(t)$ using Hilbert
transforms. The instantaneous amplitudes and phases evolve according to:
\begin{eqnarray}
\dot R_i(t)&=&2\sqrt{R_i(1+R_i^2-2R_i\cos\phi_i)}\{R_i\cos(\zeta_i-\psi_i)
- R_iB_i\cos(\zeta_i-\psi_i+\rho_i)+B_i\cos(\psi_i+\rho_i)+\xi_R\}\nonumber\\
\dot\phi_i(t)&=&2\sqrt{\frac{(1+R_i^2-2R_i\cos\phi_i)}{R_i}}\{R_i\sin(\zeta_i-\psi_i)
- R_iB_i\sin(\zeta_i-\psi_i+\rho_i)+B_i\sin(\psi_i+\rho_i)+\xi_{\phi}\}
\end{eqnarray}
where $\zeta_i(t)=\arctan(\frac{R_i\sin\phi_i}{R_i\cos\phi_i-1})$, and
$\psi_i(t)=(\zeta_i(t)-\phi_i(t))/2$  denotes the instantaneous difference
between the phase of the output of the $i$th neuron and that of the part it sends
as feedback to the presynaptic conductance of the $j$th neuron --- this feedback
is a stochastic component of the input for neuron $j$.
We have constructed the noise terms $\xi_R(t)$ and $\xi_{\phi}(t)$ from the
analytical noise signal $\xi(t)+i\nu(t)=\xi(t)+iH(\xi(t))$:
\begin{equation}
\xi_{R_i}=(1-R_i)(\xi\cos\psi_i+\nu\sin\psi_i), ~~~
\xi_{\phi_i}=(1-R_i)(\xi\sin\psi_i+\nu\cos\psi_i)
\end{equation}
$\xi_{R_i}(t)$ and $\xi_{\phi_i}(t)$ are periodically modulated
by $\psi_i$ which evolves according to the differential equation:
\begin{eqnarray}
\dot\psi_i&=&\frac{1}{\sqrt{R_i(1+R_i^2-2R_i\cos\phi_i)}}\{(R_i^2-1)[R_i\sin(\zeta_i-\psi_i)
-B_iR_i\sin(\zeta_i-\psi_i+\rho_i)+B_i\sin(\psi_i+\rho_i)\nonumber\\
&+&\xi_{\phi_i}]-R_i\sin\phi_i[R_i\cos(\zeta_i-\psi_i)-B_iR_i\cos(\zeta_i-\psi_i+\rho_i)
+ B_i\cos(\psi_i+\rho_i)+ \xi_{R_i}]\} \nonumber\\
&=&\frac{1}{2\sqrt{(1+R_i^2-2R_i\cos\phi_i)}}\big\{ (R^2-1){\sqrt R} {\dot\phi}
- \sin\phi {\dot R} \big\}
\end{eqnarray}
The instantaneous phase $\phi_i$ therefore has no deterministic time scales and
its drift and diffusion in time in the presence of noise is influenced by $B_i$
and $\rho_i$, the instantaneous values of the amplitude and phase respectively
of the input, and also by the instantaneous amplitude $R_i$ of the neuronal output.
$B_i(t)$ and $\rho_i(t)$ evolve as follows:
\begin{eqnarray}
\dot B_i(t) &=& -B_i\Big(\frac{1}{\tau_{ji}}+\frac{e^{-2\eta(1-R_j\cos\phi_j)}}
{\tau_R}\cos(2\eta R_j\sin\phi_j)\Big) + \frac{\beta_i}{\tau_{ji}}\cos\rho_i \nonumber\\
&+&\frac{1}{\tau_R}(\beta_i+\sum_j\frac{\alpha_jg_{ji}}{\tau_R})
\cos((2\eta R_j\sin\phi_j)-\rho_i)
~e^{-2\eta(1-R_j\cos\phi_j)}\nonumber\\
\dot\rho_i(t) &=& -\frac{e^{-2\eta(1-R_j\cos\phi_j)}}{\tau_R}\sin(2\eta R_j\sin\phi_j)
-\frac{\beta_i}{B_i\tau_{ji}}\sin\rho_i \nonumber\\
&+&\frac{1}{\tau_R B_i}
(\beta_i+\sum_j\frac{\alpha_jg_{ji}}{\tau_R})\sin((2\eta R_j\sin\phi_j)-\rho_i)
~e^{-2\eta(1-R_j\cos\phi_j)}
\end{eqnarray}
These show the effect of feedback on neuronal response.
It is seen that CS between the outputs of neurons 1 and 2 occurs when the
changes in the instantaneous phases and amplitudes of the two neuronal {\em inputs}
exactly match each other. In other words, CS in the inputs to the neurons is
required for the outputs to synchronize in phase, amplitude and frequency.
This observation is in general not quite obvious since the system is nonlinear,
and has a feedback mechanism which depends upon the outputs of the other neurons. \\
In Fig.(4) we present the instantaneous-phase versus instantaneous-frequency
plots of the inputs received by the two neurons. In all the numerous cases we
studied for the coupled system for $n=2$, we found that the signature of CS is
the almost identical nature of these plots for the two systems that are in
synchrony, be it with or without noise. On the other hand, the absence of CS
gets reflected in the non-identical variation between instantaneous phase and
instantaneous rates of phase-change of inputs to the two neurons, in the plots
shown in Fig.(5). This is again true for both the noiseless as well as noisy
cases. \\
The system of equations (eqn.(5)) is of the form
\begin{equation}
{\dot u}_i = f_D(u_i, s_i, u_j, t, \beta=0, \alpha_j g_{ji}=0)
+ (\beta_i + \alpha_j g_{ji},s_i(u_j, t))f_n(u_i) + \xi f_n(u_i),
\end{equation}
where
\begin{eqnarray}
f_D (u_i, s_i, \beta_i=0, \alpha_jg_{ji}=0) = 2 u_i^{3/2}(1 - u_i)^{1/2} , ~~~
f_n(u_i) = 2 u_i^{1/2}(1 - u_i)^{3/2},
\end{eqnarray}
and since feedback to each neuron through the synaptic coupling is oscillatory,
in a Fokker-Planck description of the stochastic process, the probability distribution
$P(u_1, u_2,t)$ of the neuronal ensemble will not be stationary in the $t \rightarrow \infty$
limit~\cite{hanggi}.
In certain regimes of the noise strengths and coupling constants, where competing contributions
from the drift and diffusion terms would make the noise-averaged difference in
outputs $\langle (u_1 - u_2 \rangle)$ zero, CS occurs.

Physically, CS is brought about through the following sequence of events.
Addition of a small amount of noise increases the decay time of the synaptic
conductances $s_{ij}$ gradually, and eventually, lowers their minimum to zero.
This delays the onset of the next $s_{ij}$ peaks, and hence the input $I_i$ to
each neuron at any further instant of time. Increasing the noise strength
further increases the decay time of $s_{ij}$. The periodically maximal values
of the inputs thus take longer to arrive at the neurons and this becomes
visible in the neuronal firing pattern as departures from the previous
(noiseless) values of the phase differences between the neurons, and those of
their output differences $\langle u_1-u_2\rangle$. The instantaneous values of
the phases and the rate at which they change in time, i.e., the instantaneous
frequencies of each neuron, is determined by the strengths $\alpha_{ij}g_{ij}$
of the synaptic inputs it receives and the noise strength for any given set of
$\beta, \tau_R, \tau_{ij}, \eta$. Hence for given initial conditions and
different amplitudes  for $u_1$ \& $u_2$, it would be reasonable to expect CS
to occur when the following is satisfied for the inputs to the neurons: the
variation of instantaneous values of the frequencies with instantaneous phases
of the {\em input} for neuron 1 matches with that for neuron 2. This results in
the instantaneous values $\phi_1$ and ${\dot\phi}_1$ of neuron 1 changing in
step with $\phi_2$ and ${\dot\phi}_2$ respectively of neuron 2. This forces the
amplitudes of neurons 1 \& 2 to become identical with each other, since
otherwise both conditions $\phi_1-\phi_2$ and ${\dot\phi}_1={\dot\phi}_2$
cannot be simultaneously maintained. Thus CS results.

A striking feature of all these plots is their strange, flame-like
structure. The flame shape is reminiscent of canards that are typically associated
with systems exhibiting relaxation oscillations.\\
Indeed, from the common factor occurring in the inverse square root on the
right hand side of eqn.(8), it is apparent that $\psi(t)$ does evolve on a time
scale different from that for $R(t), \phi(t), B(t)$ \& $\rho(t)$ in eqns.(6) \&
(9), though it is a dynamically varying time scale, determined also by the
stochasticity of the system. The separation of time scales is in fact
manifested in the time series for the neuronal inputs which show relaxation
oscillations (Figs.(4) \& (5)). This gives the input instantaneous
phase-frequency curve its characteristic shape whenever noise is introduced.
This is indicative of some order emerging in the phase. In fact, noise-induced
phase synchronization can be demonstrated even in the IE system by interpreting
the instantaneous phase differences $\Phi=\phi_1-\phi_2$ in a statistical sense
as in~\cite{zhou}. The distribution $P(\Phi)$, of cyclic instantaneous phase
differences, $\Phi\in[-\pi,\pi]$ is shown in Fig.(6). Preferred phase
differences between the 2 neurons manifest as peaks in $\Phi$ which become
sharper and taller with increasing noise-strengths -- a clear indication of
noise-induced phase coherence.
\section{Noise-induced CS \& Partial Synchronization in EE \& IE systems}
In Fig.(7a) we have plotted the synchronization error $\langle |u_1 - u_2|
\rangle$ for two coupled identical (EE) neurons and for coupled non-identical
(IE) neurons, as a function of the coupling strength for different
noise-strengths. We have considered here the special case of $\beta_1 = \beta_2
= 0$, and each neuron receives inputs only from the other neurons; {\it i.e.},
we look at the effect of feedback. Results and analysis for non-zero $\beta_i$
are presented elsewhere~\cite{malik}. We see that although CS is expected
between identical oscillators, it does not happen for the noiseless
(deterministic) case. Increasing noise-strength brings down $\langle |u_1 -
u_2| \rangle$ and indeed takes it down to zero for certain ranges of the
coupling constant.

For the identical EE coupled neurons, when feedback constitutes the only input (there being
no other explicit input), as in Figs.(7a,7b), then beyond a maximal critical coupling constant
strength $g_c$, the system gets de-synchronized, with $\langle|u_1 - u_2|\rangle \ne 0$.
For $g > g_c$, there then exists a regime in the de-synchronized system where noise strength
$\sigma$ still plays a role in determining the output.
In this transition regime, for $g > g_c$, we find that the difference in outputs
$\langle|u_1 - u_2|\rangle_{trans}$ depends upon $g$ through the expression
\begin{equation}
\langle|u_1 - u_2|\rangle_{trans} = (g - g_c)^{1/4} - \nu_{\sigma},
\end{equation}
where $\nu_{\sigma}$ is a constant that depends upon the noise strength $\sigma$.\\
\noindent
The critical coupling strength $g_c$ depends upon the noise strength as well and we
find that it varies as
\begin{equation}
g_c \sim a \sqrt{\sigma},
\end{equation}
where $a=1.1$.
At very high $g$ the system gets locked to a partially synchronized state, with
$\langle|u_1 - u_2|\rangle$ approaching a constant value, wherein noise strength no longer
influences the difference in the outputs of the neurons. In Fig.(7a), this constant value
approaches 0.5. For the curves in Fig.(7a,b), for the entire regime following the beginning
of desynchronization, we find a functional dependence on the coupling constant $g$
given by an equation of the form
\begin{equation}
\langle|u_1 - u_2|\rangle_{\sigma} \sim a(\sigma) - b(\sigma) g - \frac{{g_c}^4}{g^3},
~~~~~~~~~~({\rm for} ~~g > g_c),
\end{equation}
where $a(\sigma)$ and $b(\sigma)$ depend on the noise-strength through
\begin{equation}
a(\sigma) = \frac{0.59}{\sigma^{0.13}},~~~~ b(\sigma) = \frac{0.0181}{\sigma^{0.53}},
\end{equation}
so that
\begin{equation}
\langle|u_1 - u_2|\rangle_{\sigma} \sim \frac{0.59}{\sigma^{0.13}} - \frac{0.0181}{\sigma^{0.53}}g
- \frac{1.4641}{g^3}\sigma^{2.01} ~~~~~~~~~~({\rm for} ~~g > g_c).
\end{equation}
This expression is plotted on the numerical data points in Fig.(7b). A rigorous
theoretical treatment of the system needs to be done in future studies to
establish these relations for the synchronization and desynchronization
transitions through a Fokker-Planck approach. \\
In the IE case (Fig.(7c)), we see that noise-induced CS does not occur for two
coupled neurons; however, there is partial synchronization (by this we mean
that $\langle |u_1 - u_2| \rangle = {\rm constant} \ne 0$) since $\langle |u_1
- u_2| \rangle$ gets locked to a finite, non-zero value ($\sim 0.6$) for large
$g$. Further, increasing noise-strength increases rather than decreases
$\langle |u_1 - u_2| \rangle$, in the region before partial synchronization, in
contrast to the observation for the EE case. We were unable to achieve CS in
the IE case for two coupled neurons, even in the presence of noise.

However, for an ensemble of 200 coupled theta neurons of which 150 neurons are
excitatory and 50 inhibitory, we obtain very different results. In this
simulation shown in Fig.(8), each neuron receives the same input $\beta_i=0.1$
and there is all to all random coupling with the coupling strengths varying
between $0.05$ and $0.1$ and with different initial conditions. The excitatory
neurons are shown in red while the inhibitory ones are in green. On introducing
Gaussian white noise of strength $\sigma=5.0$ into the system we observe
synchronous phenomena emerging between the excitatory and inhibitory neurons.
The interesting thing to note here is that not only do most of the inhibitory
neurons fire in synchrony with other inhibitory neurons, but that also most of
them are synchronized with the excitatory neurons. This kind of noise-induced
near-CS in coupled type-I neurons with random non-zero coupling strengths has
not, to our knowledge, been reported previously in the literature. A detailed
study of this situation is under way to explain the observed synchrony for
$n=200$ and will be reported elsewhere: it is beyond the scope of the present
work. Interestingly, in the literature, spatiotemporal synchronization has been 
shown to occur in networks of coupled chaotic maps with varying degrees of 
randomness in the coupling connections~\cite{sinha}. 

\section{Conclusion}
We have studied the issue of noise-induced CS in coupled type-I neurons, a
class of neurons that are especially important since mammalian neurons fall
under this category. We find that complete synchronization between any two
neurons is signalled whenever the variations of the instantaneous input phases
$\phi$ versus the instantaneous frequencies ${\dot \phi}$ of the inputs of the
neurons being studied are identical. We point out that such identical
$\phi$-${\dot \phi}$ plots of the neuronal inputs would be a signature of CS
between the neurons. This suggests the possibility of producing completely
synchronized outputs of coupled systems having feedback mechanisms in the
presence of noise, by ensuring that the plots of the instantaneous frequency
versus instantaneous phase of the {\em inputs} to the subsystems are identical.
That this is a significant point will be appreciated when one recalls that this
condition is required in the continued presence of noise and feedback. When
monitoring or control of neurons is required to be done in a living organism in
the case where synchronized neuronal output is required at another,
inaccessible spot through a given external input, this result becomes
important.

Though CS is expected between coupled,
identical neurons, we find that CS occurs only upto a critical value $g_c$
of the coupling constant $g$ for a given noise strength beyond which the
system desynchronizes again, and then for large $g$ gets locked to a
partially-synchronized state. We find that the critical coupling strength
$g_c$ depends upon the noise-strength through a power law. For $g$ greater
than the critical value $g_c$, from the transition regime through upto the
onset of partial synchronization, we find a functional dependence of the
noise-averaged output difference on $g$ and the noise strength $\sigma$,
given by eqn.(14). For a larger ensemble of 200 neurons, we find  unexpectedly
that non-identical neurons can show near-complete synchronization.

Since type-I neurons model axonal excitability patterns in mammals, the results
presented here would be useful in the study of synchronous mechanisms
underlying the neural code. As an immediate application, we believe our results
would be useful in explaining the experimental observations reported on cat \&
awake monkey visual cortex ~\cite{singer,kreiter,gerstner,singer2,ritz} which
show synchronization of neuronal activity with a single stimulus, and which
disappears when activated by different, independent stimuli. Since a single
stimulus would correspond to neuronal inputs of identical amplitudes,
instantaneous phase and frequency, this is actually the same scenario that we
have found for CS to occur. It is also likewise clear why CS was experimentally
observed to vanish on the activation of different stimuli, since the necessary
conditions of identical input amplitude, phase and frequency are no longer met.

\newpage
\noindent
\begin{center}
\includegraphics[height=4.5cm,width=8.5cm,angle=0]{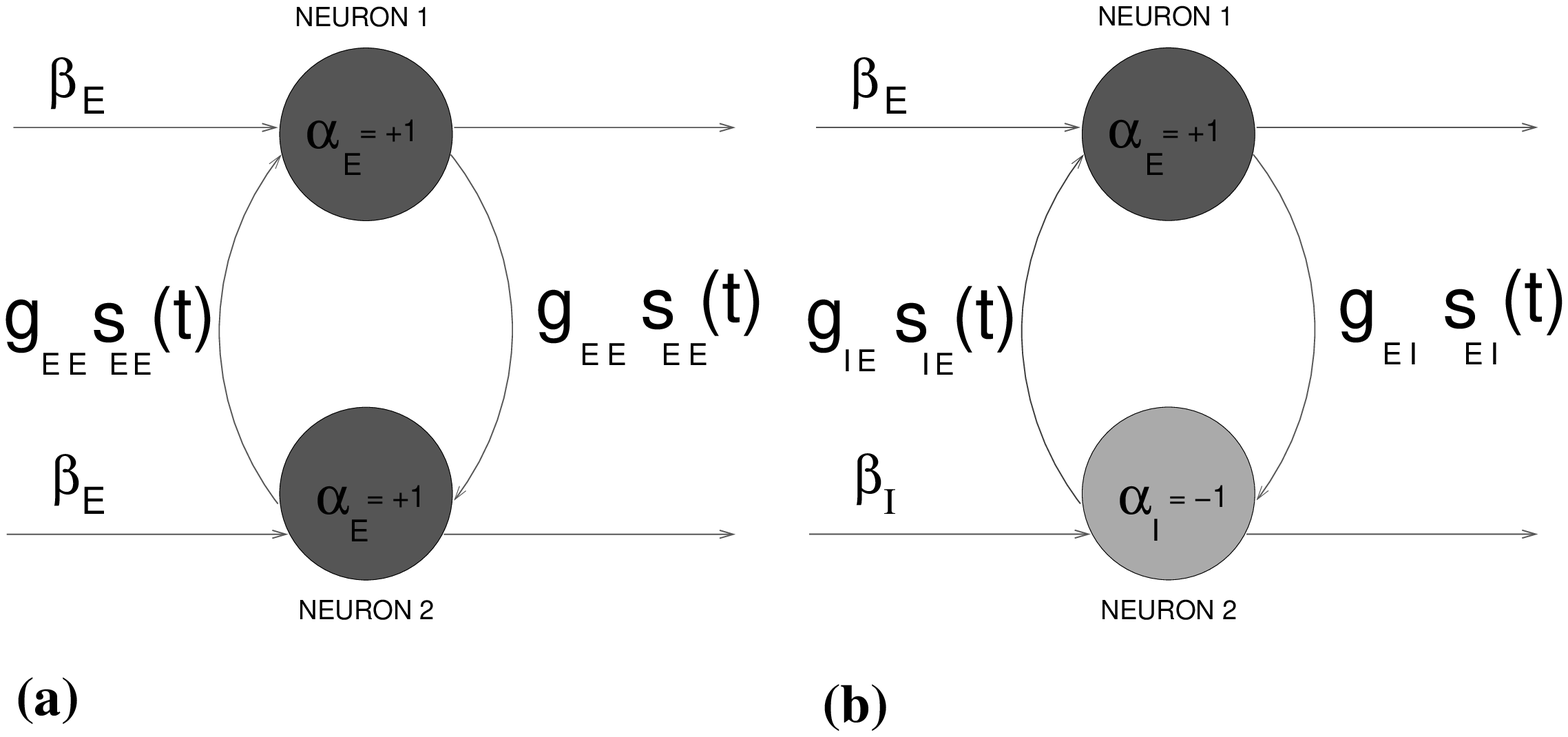}
\end{center}
{\bf Figure 1.} Bidirectional coupling between two theta neurons.
Left: E-E coupling, Right: I-E coupling.\\
\newpage
\begin{center}
\includegraphics[height=10cm,width=6.5cm,angle=0]{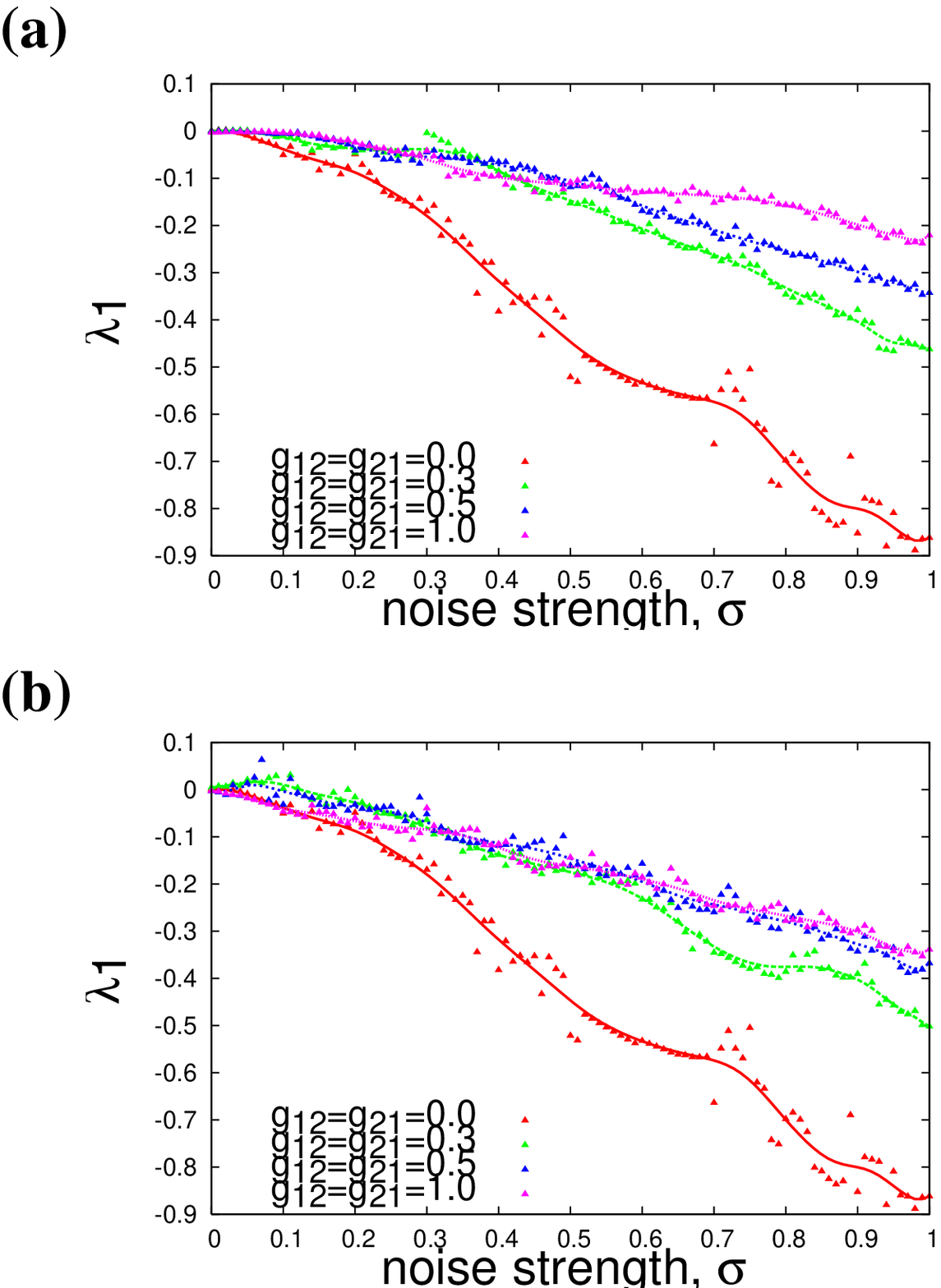}
\vspace*{0.5cm}
\end{center}
{\bf Figure 2.} Lyapunov exponent in the presence of noise for:
(at top) EE coupling and (at bottom) IE coupling. In both cases, the parameters
which we used are $\beta_1=\beta_2=0.1, \tau_{12}=\tau_{21}=2.0, \tau_R=0.1, \eta=5.0$.\\
\begin{center}
\includegraphics[height=5.5cm,width=4.6cm,angle=270]{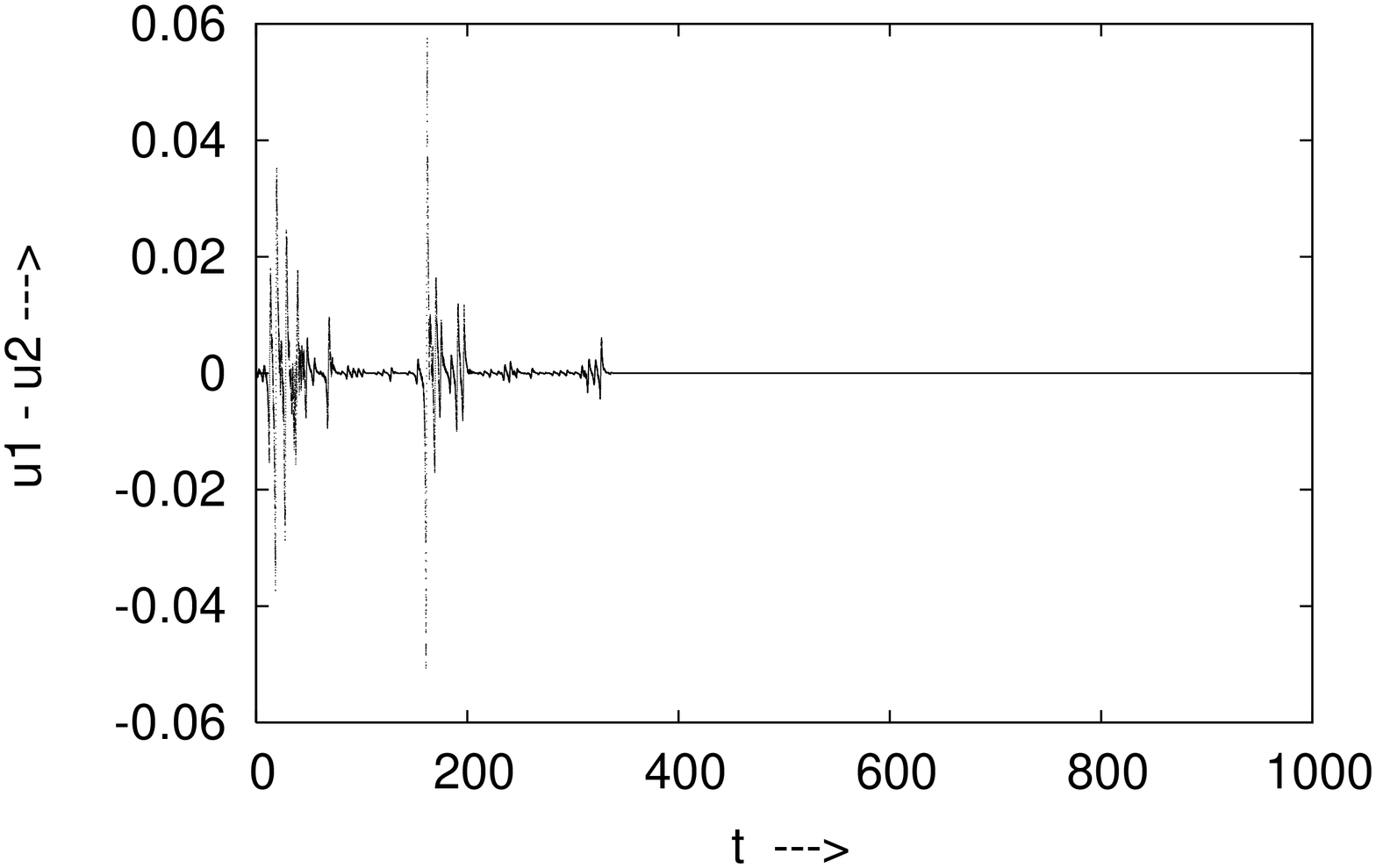}
\end{center}
{\bf Figure 3.} Windows of zero synchronization error interspersed with
intermittent bursts of firing in coupled identical EE neurons: $g_{12} = g_{21}
= 0.3$, $\tau_{12} = \tau_{21} = 2.0$,
$\tau_R = 0.1$, $\beta_1 = \beta_2 = 0.1$, $\sigma = 0.2655$. 
\newpage
\begin{center}
\includegraphics[height=6.9cm,width=12.3cm,angle=0]{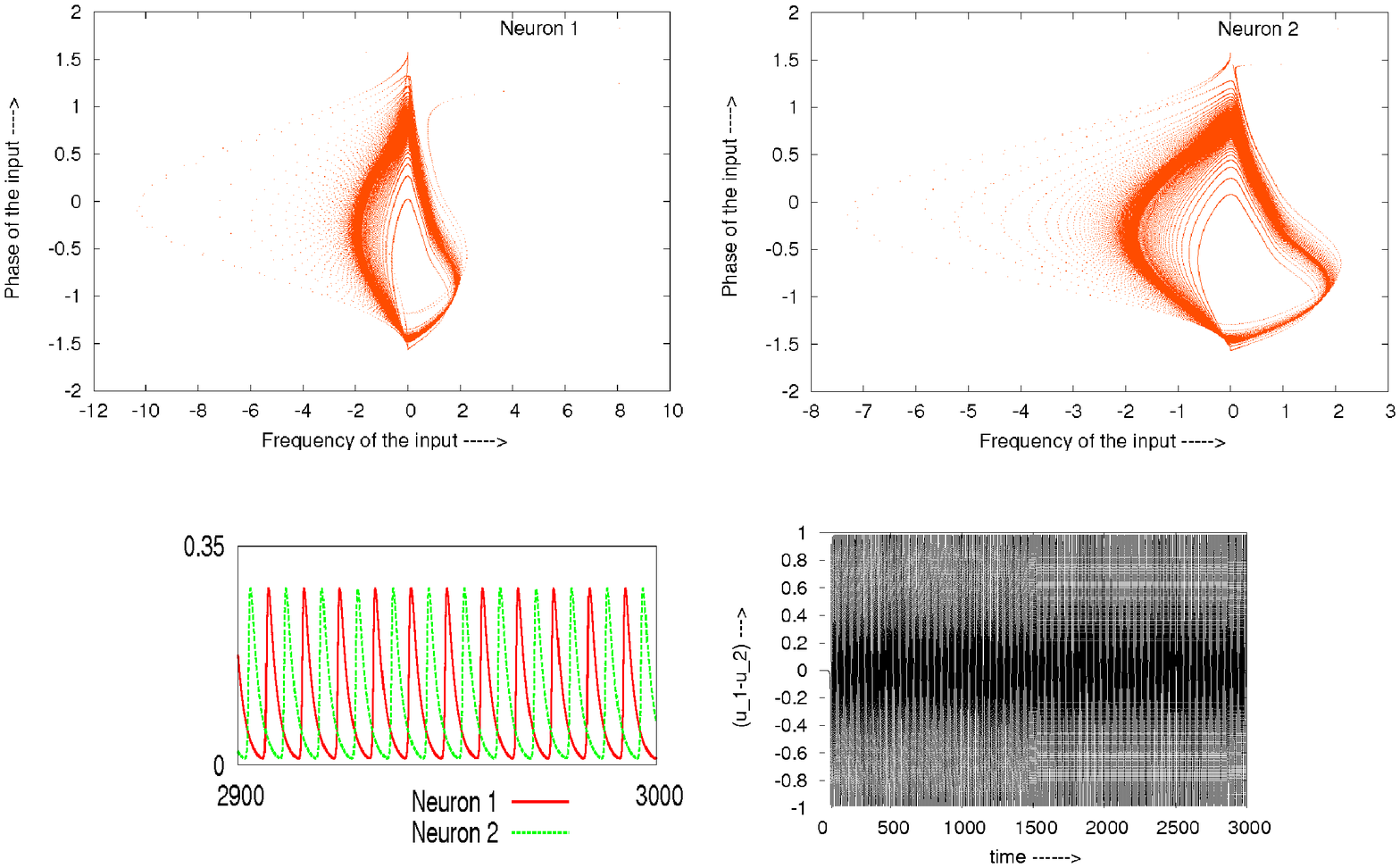}
\end{center}
{\bf a}\\
\begin{center}
\includegraphics[height=6.9cm,width=12.3cm,angle=0]{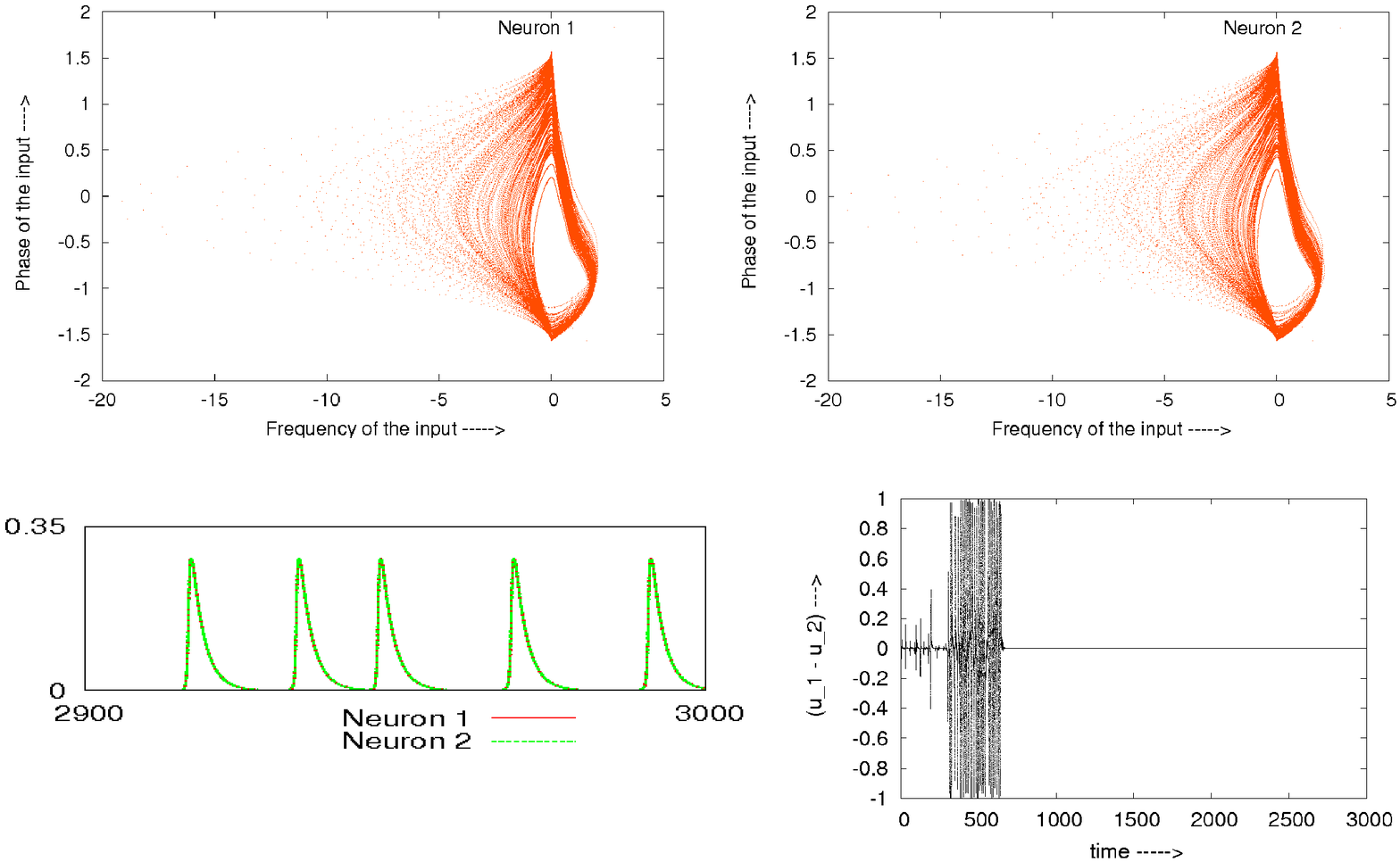}
\end{center}
{\bf b}\\ 
{\bf Figure 4.} Instantaneous phase-frequency ``flame" plots of
stochastic {\bf input} $I_i=\beta_i+\sum_{j=1}^N\alpha_j g_{ji}s_{ji}$ to
neuron $i$, for EE coupling, for {\bf (a)}~$\sigma = 0$ (top) and {\bf (b)}~$\sigma=0.09$ (bottom).
In these plots, $\beta_1=\beta_2=0.0$, showing the effect of feedback. The
extended panels below each set of flame plots show the corresponding
time-series for neuronal input (at left) and the difference ($u_1-u_2$) in
neuronal outputs (at right).
Parameter values are:  $g_{12}=g_{21}=0.3$, $\tau_{12}=\tau_{21}=2.0$, $\tau_R=0.1$.\\
\newpage
\begin{center}
\includegraphics[height=13.2cm,width=14.2cm,angle=0]{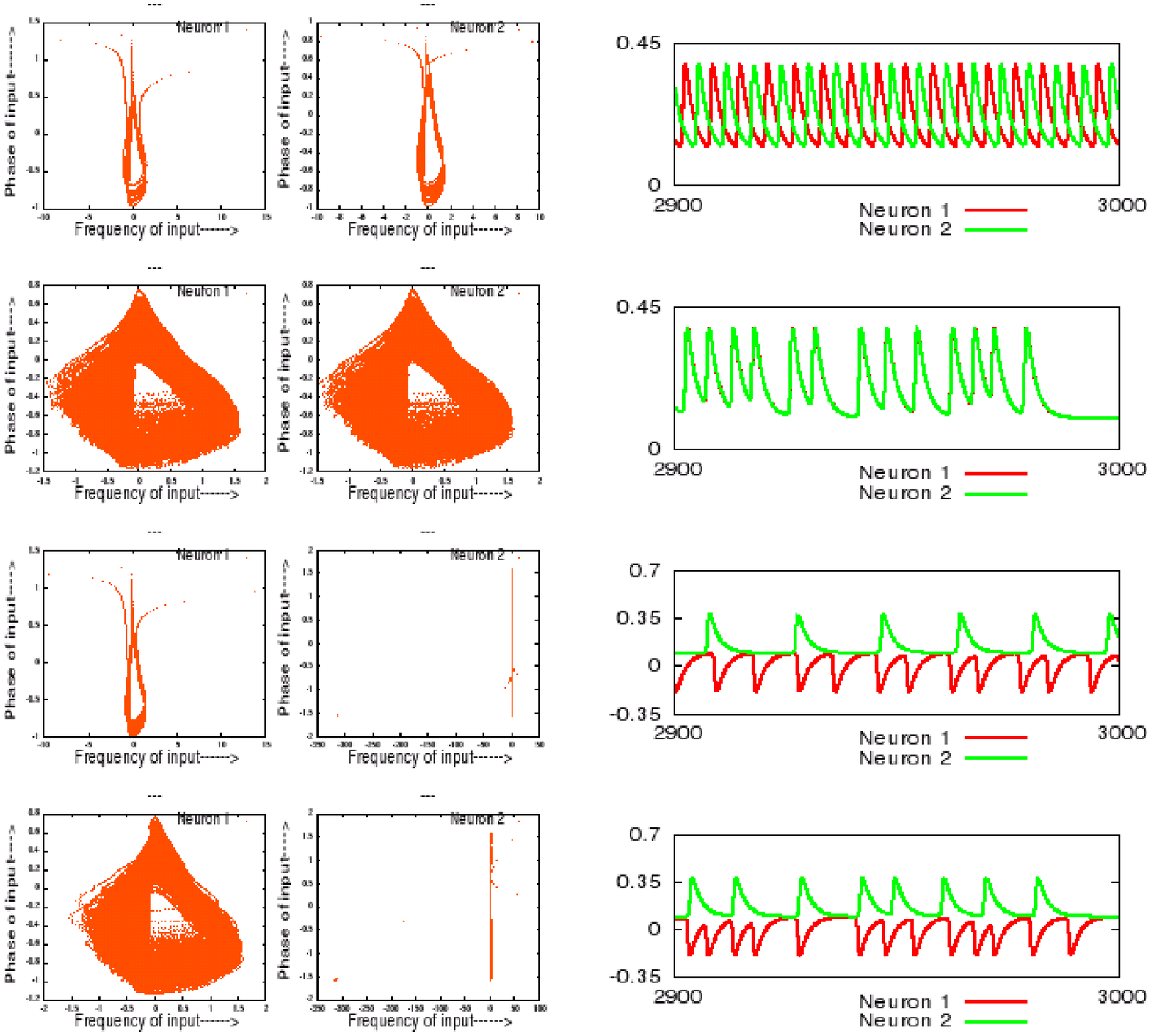}
\end{center}
{\bf Figure 5.} More instantaneous input phase-frequency
``flame" plots with corresponding time-series of neuronal input. row 1: EE,
$\sigma=0.0$; row 2: EE, $\sigma=0.301$; row 3: IE, $\sigma=0.0$; row 4: IE,
$\sigma=0.301$. Parameters used are:$g_{12}=g_{21}=0.3$,
$\tau_{12}=\tau_{21}=2.0$, $\beta_1=\beta_2=0.1$, $\tau_R=0.1$ .\\
\newpage
\begin{center}
\includegraphics[height=8.5cm,width=7.1cm,angle=270]{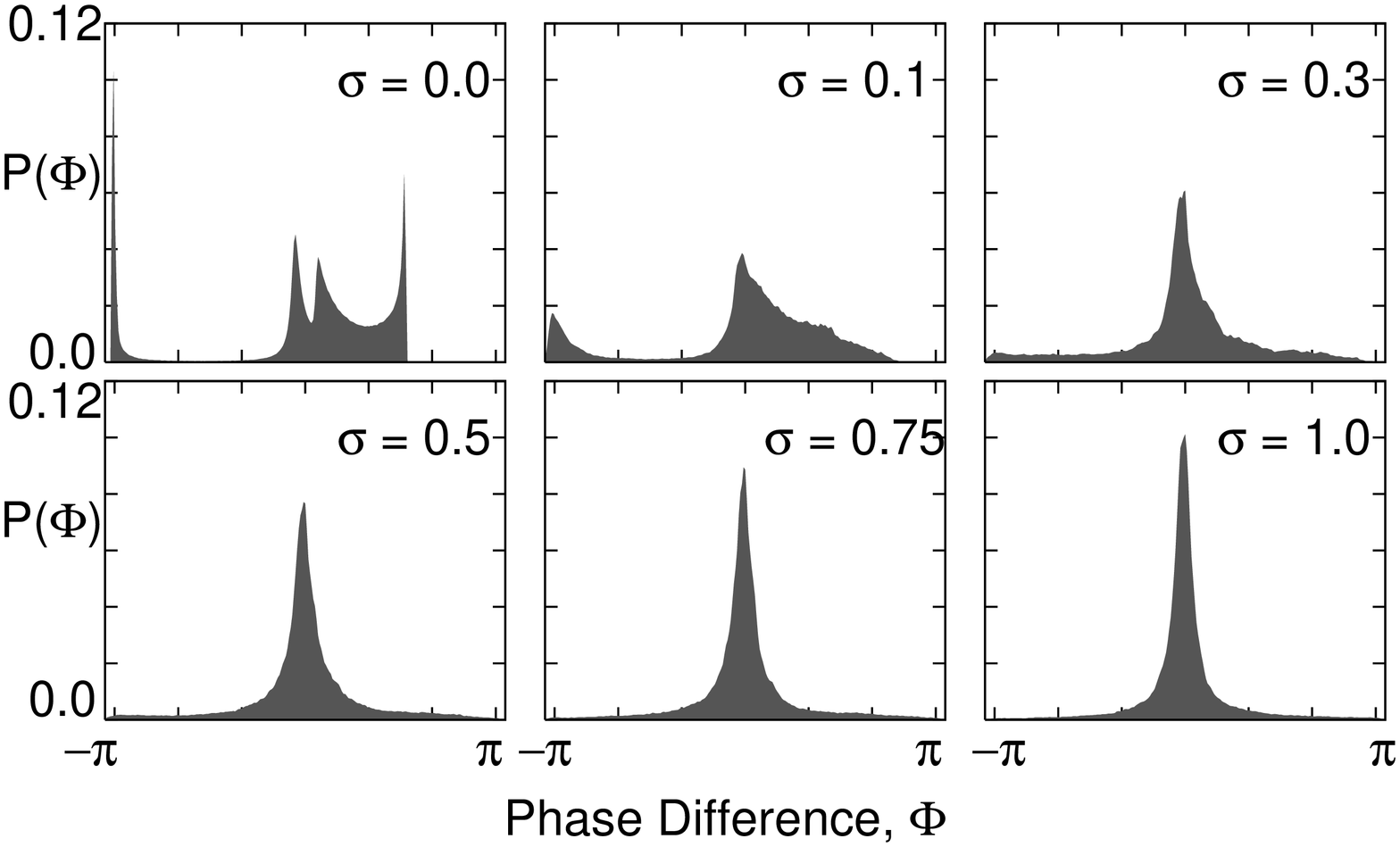}
\end{center}
{\bf Figure 6.} Distribution of instantaneous phase-differences for neuron pair with IE coupling,
at different noise strengths. Here, $g_{12}=g_{21}=0.3, \tau_{12}=\tau_{21}=2.0, \tau_R=0.1,
\eta=5.0, \beta_1=\beta_2=0.0, \alpha_1=+1, \alpha_2=-1$.\\
\newpage
\begin{center}
\includegraphics[height=8.5cm,width=16cm,angle=270]{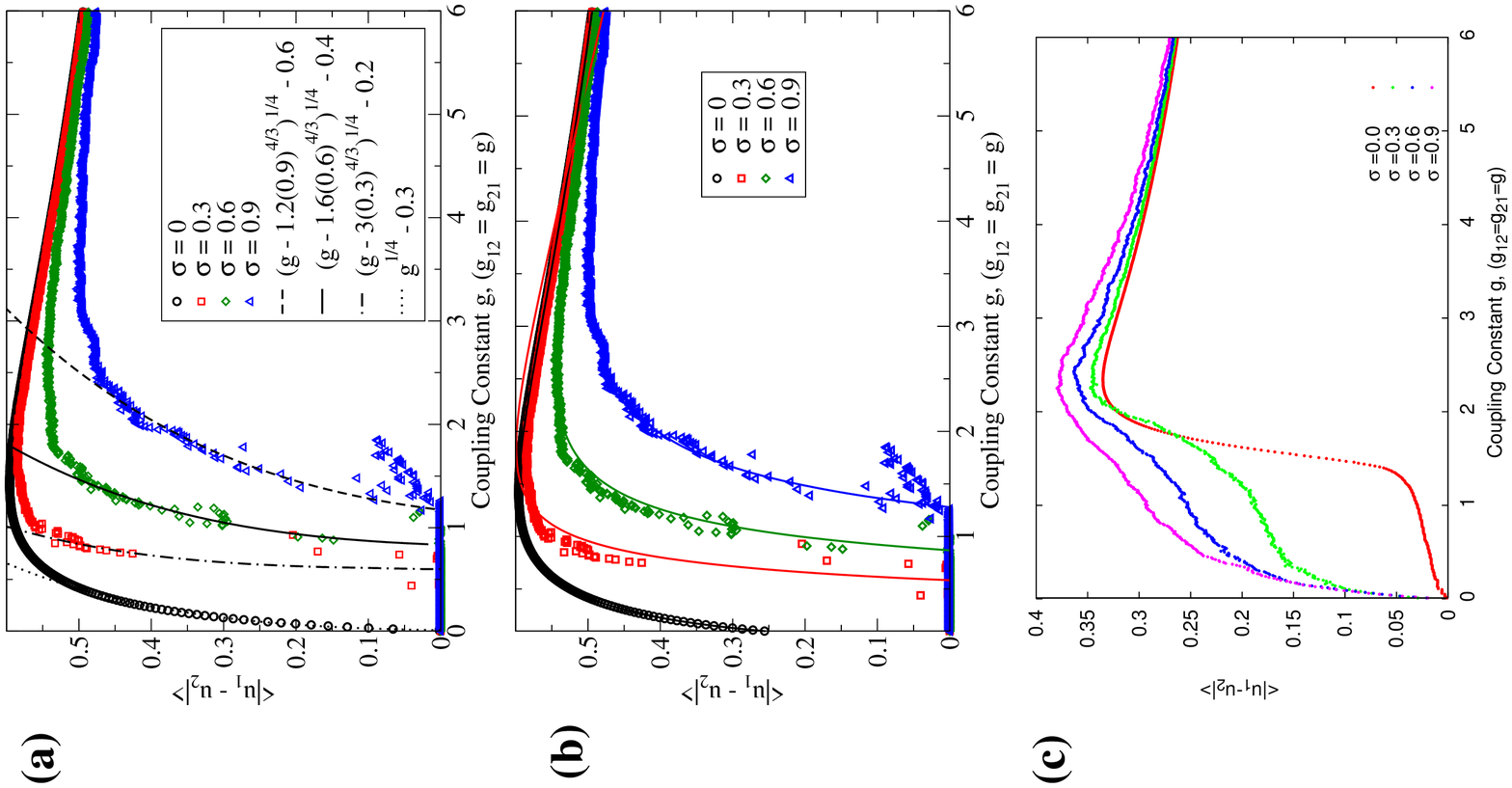}
\vspace*{0.5cm}
\end{center}
{\bf Figure 7.} 
Noise-induced synchronization in coupled Type-I neurons (a) EE case; transition from synchronized
to partially synchronized state when $g > g_c$ (b) EE case, as in (a); solid lines correspond to Eqn.(16).
(c) IE case; noise-induced CS is absent but there
is partial synchronization as the system gets locked to $\langle|u_1 - u_2|\rangle \sim 0.6$ for
large $g$. For both (a) \& (b): $\tau_{12} = \tau_{21} = 2.0$, $\beta_1 = \beta_2 = 0.0$,
$\tau_{R} = 0.1$, initial conditions $\theta_1 = 0.0$, $\theta_2 =0.01$, $s_{12} = s_{21}=0.0$\\
\newpage
\begin{center}
\includegraphics[height=8cm,width=12cm,angle=270]{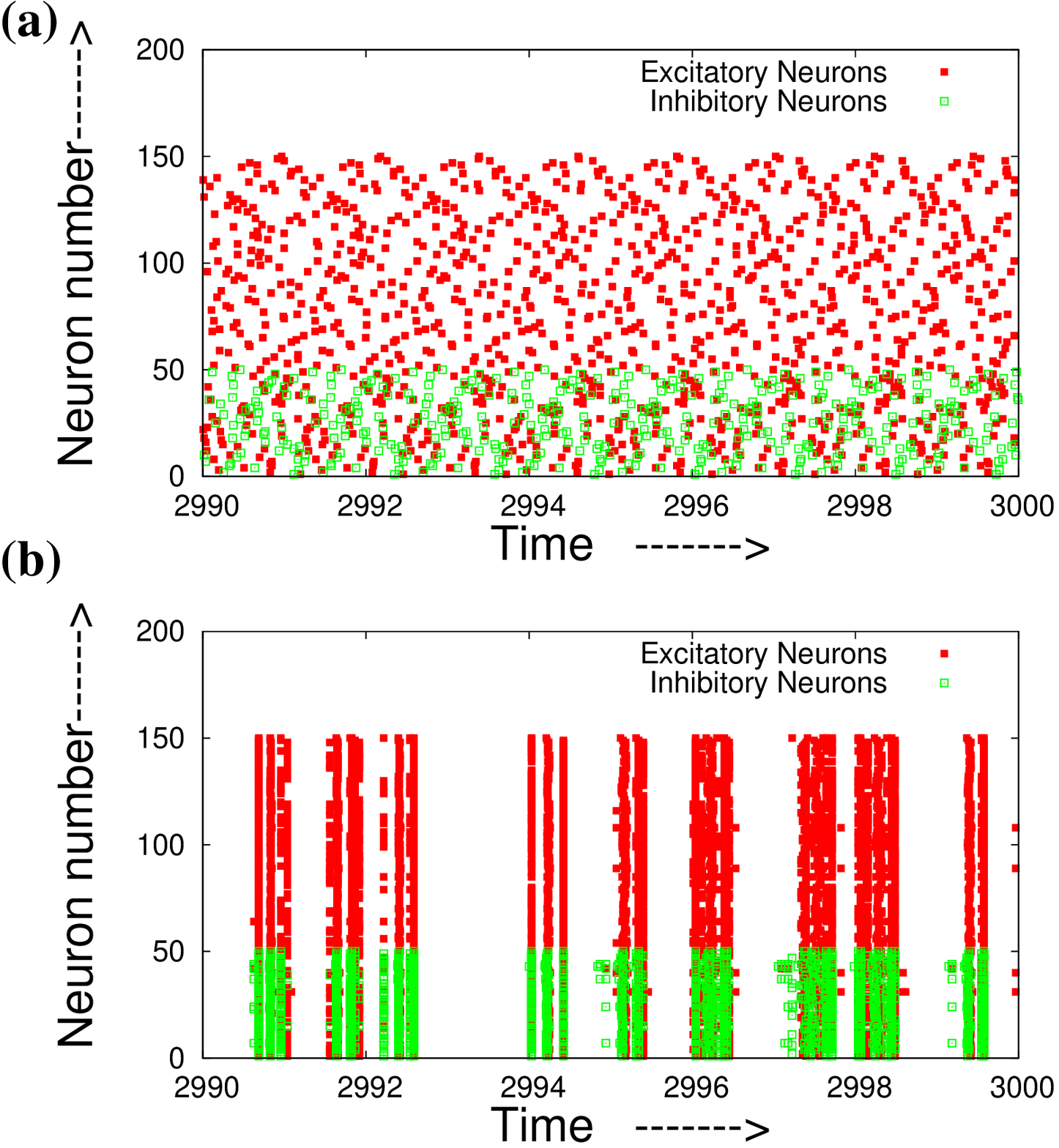}
\end{center}
{\bf Figure 8.} Simulation of 200 bidirectionally-coupled theta neurons
(150 EE (in red) and 50 IE (in green)) with random all-to-all couplings;
top: $\sigma=0$, bottom: $\sigma=5.0$.

\begin{thebibliography}{20}
\expandafter\ifx\csname natexlab\endcsname\relax\def\natexlab#1{#1}\fi
\expandafter\ifx\csname bibnamefont\endcsname\relax
  \def\bibnamefont#1{#1}\fi
\expandafter\ifx\csname bibfnamefont\endcsname\relax
  \def\bibfnamefont#1{#1}\fi
\expandafter\ifx\csname citenamefont\endcsname\relax
  \def\citenamefont#1{#1}\fi
\expandafter\ifx\csname url\endcsname\relax
  \def\url#1{\texttt{#1}}\fi
\expandafter\ifx\csname urlprefix\endcsname\relax\def\urlprefix{URL }\fi
\providecommand{\bibinfo}[2]{#2}
\providecommand{\eprint}[2][]{\url{#2}}

\bibitem[{\citenamefont{Pecora and Carroll}(1990)}]{pecora}
\bibinfo{author}{\bibfnamefont{L.M.}~\bibnamefont{Pecora}} \bibnamefont{and}
  \bibinfo{author}{\bibfnamefont{T.L.} \bibnamefont{Carroll}},
  \bibinfo{journal}{Phys.Rev.Lett.} \textbf{\bibinfo{volume}{64}},
  \bibinfo{pages}{821} (\bibinfo{year}{1990}).

\bibitem[{\citenamefont{Hansel and Sompolinsky}(1992)}]{hanselsompolinsky}
\bibinfo{author}{\bibfnamefont{D.}~\bibnamefont{Hansel}} \bibnamefont{and}
  \bibinfo{author}{\bibfnamefont{H.} \bibnamefont{Sompolinsky}},
  \bibinfo{journal}{Phys.Rev.Lett.} \textbf{\bibinfo{volume}{68}},
  \bibinfo{pages}{718} (\bibinfo{year}{1992}).

\bibitem[{\citenamefont{Wang, Perez and Cerdeira}(1993)}]{cerdeira}
\bibinfo{author}{\bibfnamefont{W.}~\bibnamefont{Wang}}, 
  \bibinfo{author}{\bibfnamefont{G.} \bibnamefont{Perez}}, \bibnamefont{and}
  \bibinfo{author}{\bibfnamefont{H.A.} \bibnamefont{Cerdeira}},
  \bibinfo{journal}{Phys.Rev. E} \textbf{\bibinfo{volume}{47}},
  \bibinfo{pages}{2893} (\bibinfo{year}{1993}).

\bibitem[{\citenamefont{Vreeswijk, Abbott and Ermentrout}(1994)}]{vreeswijk}
\bibinfo{author}{\bibfnamefont{C.}~\bibnamefont{van Vreeswijk}}, 
\bibinfo{author}{\bibfnamefont{L.F.}~\bibnamefont{Abbott}} \bibnamefont{and}
  \bibinfo{author}{\bibfnamefont{G.B.} \bibnamefont{Ermentrout}},
  \bibinfo{journal}{J.Comput.Neurosci.} \textbf{\bibinfo{volume}{1}},
  \bibinfo{pages}{313} (\bibinfo{year}{1994}).

\bibitem[{\citenamefont{Hansel et~al.}(1995)\citenamefont{Hansel, Mato, and
  Meunier}}]{hansel}
\bibinfo{author}{\bibfnamefont{D.}~\bibnamefont{Hansel}},
  \bibinfo{author}{\bibfnamefont{G.}~\bibnamefont{Mato}}, \bibnamefont{and}
  \bibinfo{author}{\bibfnamefont{C.}~\bibnamefont{Meunier}},
  \bibinfo{journal}{Neural Comput.} \textbf{\bibinfo{volume}{7}},
  \bibinfo{pages}{307} (\bibinfo{year}{1995}).

\bibitem[{\citenamefont{Ermentrout}(1996)}]{ermentrout}
\bibinfo{author}{\bibfnamefont{B.}~\bibnamefont{Ermentrout}},
  \bibinfo{journal}{Neural Comput.} \textbf{\bibinfo{volume}{8}},
  \bibinfo{pages}{979} (\bibinfo{year}{1996}).

\bibitem[{\citenamefont{B{\"o}rgers and Kopell}(2003)}]{borgerskopell}
\bibinfo{author}{\bibfnamefont{C.}~\bibnamefont{B{\"o}rgers}} \bibnamefont{and}
  \bibinfo{author}{\bibfnamefont{N.}~\bibnamefont{Kopell}},
  \bibinfo{journal}{Neural Comput.} \textbf{\bibinfo{volume}{15}},
  \bibinfo{pages}{509} (\bibinfo{year}{2003}).

\bibitem[{\citenamefont{B{\"o}rgers and Kopell}(2005)}]{borgerskopell2}
\bibinfo{author}{\bibfnamefont{C.}~\bibnamefont{B{\"o}rgers}} \bibnamefont{and}
  \bibinfo{author}{\bibfnamefont{N.}~\bibnamefont{Kopell}},
  \bibinfo{journal}{Neural Comput.} \textbf{\bibinfo{volume}{17}},
  \bibinfo{pages}{557} (\bibinfo{year}{2005}).

\bibitem[{\citenamefont{Izhikevich}(1999)}]{izhikevich}
\bibinfo{author}{\bibfnamefont{E.}~\bibnamefont{Izhikevich}},
  \bibinfo{journal}{IEEE Trans. Neural Networks} \textbf{\bibinfo{volume}{10}},
  \bibinfo{pages}{499} (\bibinfo{year}{1999}).

\bibitem[{\citenamefont{Hassan, Zhang, Cerdeira and Ibiyinka}(2003)}]{cerdeira2}
\bibinfo{author}{\bibfnamefont{F.E-N.}~\bibnamefont{Hassan}}, 
\bibinfo{author}{\bibfnamefont{Y.}~\bibnamefont{Zhang}}, 
  \bibinfo{author}{\bibfnamefont{H.A.} \bibnamefont{Cerdeira}}, \bibnamefont{and}
 \bibinfo{author}{\bibfnamefont{A.F.} \bibnamefont{Ibiyinka}},
  \bibinfo{journal}{Chaos} \textbf{\bibinfo{volume}{13}},
  \bibinfo{pages}{1216} (\bibinfo{year}{2003}).

\bibitem[{\citenamefont{Hassan, Paulsamy, Fernando and Cerdeira}(2009)}]{cerdeira3}
\bibinfo{author}{\bibfnamefont{F.E-N.}~\bibnamefont{Hassan}}, 
\bibinfo{author}{\bibfnamefont{M.}~\bibnamefont{Paulsamy}}, 
  \bibinfo{author}{\bibfnamefont{F.F.} \bibnamefont{Fernando}}, \bibnamefont{and}
\bibinfo{author}{\bibfnamefont{H.A.} \bibnamefont{Cerdeira}},
  \bibinfo{journal}{Chaos} \textbf{\bibinfo{volume}{19}},
  \bibinfo{pages}{013103} (\bibinfo{year}{2009}).

\bibitem[{\citenamefont{Sudeshna Sinha}(2002)}]{sinha}
\bibinfo{author}{\bibfnamefont{S.}~\bibnamefont{Sinha}},
  \bibinfo{journal}{Phys. Rev. E} \textbf{\bibinfo{volume}{66}},
  \bibinfo{pages}{016209} (\bibinfo{year}{2002}).

\bibitem[{\citenamefont{Jampa and Sinha}(2007)}]{jampa}
\bibinfo{author}{\bibfnamefont{M.P.K.}~\bibnamefont{Jampa}}, 
\bibinfo{author}{\bibfnamefont{A.R.}~\bibnamefont{Sonawane}},
 \bibinfo{author}{\bibfnamefont{P.M.}~\bibnamefont{Gade}} \bibnamefont{and}
  \bibinfo{author}{\bibfnamefont{S.}~\bibnamefont{Sinha}},
  \bibinfo{journal}{Phys. Rev. E} \textbf{\bibinfo{volume}{75}},
  \bibinfo{pages}{026215} (\bibinfo{year}{2007}).

\bibitem[{\citenamefont{Pikovsky et~al.}(2001)\citenamefont{Pikovsky,
  Rosenblum, and Kurths}}]{pikovsky}
\bibinfo{author}{\bibfnamefont{A.}~\bibnamefont{Pikovsky}},
  \bibinfo{author}{\bibfnamefont{M.}~\bibnamefont{Rosenblum}},
  \bibnamefont{and} \bibinfo{author}{\bibfnamefont{J.}~\bibnamefont{Kurths}},
  \emph{\bibinfo{title}{Synchronization: A Universal Concept in Nonlinear
  Sciences}} (\bibinfo{publisher}{Cambridge University Press},
  \bibinfo{address}{Cambridge}, \bibinfo{year}{2001}).

\bibitem[{\citenamefont{Tuckwell}(1989)\citenamefont{Tuckwell}}]{tuckwell}
\bibinfo{author}{\bibfnamefont{H.C.}~\bibnamefont{Tuckwell}},
  \emph{\bibinfo{title}{Stochastic Processes in Neurosciences}} 
  (\bibinfo{publisher}{SIAM},
  \bibinfo{address}{Philadelphia}, \bibinfo{year}{1989}).

\bibitem[{\citenamefont{Hodgkin}(1952)}]{hodgkin}
\bibinfo{author}{\bibfnamefont{A.~L.} \bibnamefont{Hodgkin}}, 
\bibinfo{author}{\bibfnamefont{A.~F.} \bibnamefont{Huxley}}, 
  \bibinfo{journal}{J. Physiol.(London)} \textbf{\bibinfo{volume}{117}},
  \bibinfo{pages}{500} (\bibinfo{year}{1952}).

\bibitem[{\citenamefont{Rinzel and Ermentrout}(1989)}]{rinzel}
\bibinfo{author}{\bibfnamefont{J.~R.} \bibnamefont{Rinzel}} \bibnamefont{and}
  \bibinfo{author}{\bibfnamefont{G.~B.} \bibnamefont{Ermentrout}}, in
  \emph{\bibinfo{booktitle}{Methods of Neuronal Modeling}}, edited by
  \bibinfo{editor}{\bibfnamefont{C.}~\bibnamefont{Koch}} \bibnamefont{and}
  \bibinfo{editor}{\bibfnamefont{I.}~\bibnamefont{Segev}}
  (\bibinfo{publisher}{MIT Press}, \bibinfo{address}{Cambridge, MA, USA.},
  \bibinfo{year}{1989}), pp. \bibinfo{pages}{135--169}.

\bibitem[{\citenamefont{Koch}(1999)\citenamefont{Koch}}]{koch}
\bibinfo{author}{\bibfnamefont{C.}~\bibnamefont{Koch}},
  \emph{\bibinfo{title}{Biophysics of Computation: Information processing
  in single neurons}} (\bibinfo{publisher}{Oxford University Press},
  \bibinfo{address}{NY}, \bibinfo{year}{1999}).

\bibitem[{\citenamefont{Lim and Kim}(2007)}]{lim}
\bibinfo{author}{\bibfnamefont{W.}~\bibnamefont{Lim}} \bibnamefont{and}
  \bibinfo{author}{\bibfnamefont{S.-Y.} \bibnamefont{Kim}},
  \bibinfo{journal}{J. Korean Physical Soc.} \textbf{\bibinfo{volume}{50}},
  \bibinfo{pages}{219} (\bibinfo{year}{2007}).

\bibitem[{\citenamefont{Anishchenko et~al.}(2002)\citenamefont{Anishchenko,
  Astakhov, Neiman, Vadisova, and Schimansky-Geier}}]{anishchenko}
\bibinfo{author}{\bibfnamefont{V.~S.} \bibnamefont{Anishchenko}},
  \bibinfo{author}{\bibfnamefont{V.~V.} \bibnamefont{Astakhov}},
  \bibinfo{author}{\bibfnamefont{A.~B.} \bibnamefont{Neiman}},
  \bibinfo{author}{\bibfnamefont{T.~E.} \bibnamefont{Vadisova}},
  \bibnamefont{and}
  \bibinfo{author}{\bibfnamefont{L.}~\bibnamefont{Schimansky-Geier}},
  \emph{\bibinfo{title}{Nonlinear Dynamics of Chaotic and Stochastic Systems}}
  (\bibinfo{publisher}{Springer-Verlag}, \bibinfo{address}{Berlin},
  \bibinfo{year}{2002}).

\bibitem[{\citenamefont{Zhou and Kurths}(2002)}]{zhou}
\bibinfo{author}{\bibfnamefont{C.}~\bibnamefont{Zhou}} \bibnamefont{and}
  \bibinfo{author}{\bibfnamefont{J.}~\bibnamefont{Kurths}},
  \bibinfo{journal}{Phys. Rev. Lett.} \textbf{\bibinfo{volume}{88}},
  \bibinfo{pages}{230602} (\bibinfo{year}{2002}).

\bibitem[{\citenamefont{Zhou and Kurths}(2003)}]{zhou2}
\bibinfo{author}{\bibfnamefont{C.}~\bibnamefont{Zhou}} \bibnamefont{and}
  \bibinfo{author}{\bibfnamefont{J.}~\bibnamefont{Kurths}},
  \bibinfo{journal}{Chaos} \textbf{\bibinfo{volume}{13}},
  \bibinfo{pages}{401} (\bibinfo{year}{2003}).

\bibitem[{\citenamefont{Wang, Chik and Wang}(2000)}]{wang}
\bibinfo{author}{\bibfnamefont{Y.}~\bibnamefont{Wang}}, 
\bibinfo{author}{\bibfnamefont{D.T.W.}~\bibnamefont{Chik}}, \bibnamefont{and}
  \bibinfo{author}{\bibfnamefont{Z.D.}~\bibnamefont{Wang}},
  \bibinfo{journal}{Phys. Rev. E} \textbf{\bibinfo{volume}{61}},
  \bibinfo{pages}{740} (\bibinfo{year}{2000}).

\bibitem[{\citenamefont{Toral, Mirasso, Garcia and Piro}(2001)}]{toral}
\bibinfo{author}{\bibfnamefont{R.}~\bibnamefont{Toral}}, 
\bibinfo{author}{\bibfnamefont{C.R.}~\bibnamefont{Mirasso}}, 
\bibinfo{author}{\bibfnamefont{E.}~\bibnamefont{Hern\'{a}ndez-Garcia}}, \bibnamefont{and}
  \bibinfo{author}{\bibfnamefont{O.}~\bibnamefont{Piro}},
  \bibinfo{journal}{Chaos} \textbf{\bibinfo{volume}{11}},
  \bibinfo{pages}{665} (\bibinfo{year}{2001})

\bibitem[{\citenamefont{He, Shi and Stone}(2003)}]{he}
\bibinfo{author}{\bibfnamefont{D.}~\bibnamefont{He}}, 
\bibinfo{author}{\bibfnamefont{P.}~\bibnamefont{Shi}} \bibnamefont{and}
  \bibinfo{author}{\bibfnamefont{L.}~\bibnamefont{Stone}},
  \bibinfo{journal}{Phys. Rev. E} \textbf{\bibinfo{volume}{67}},
  \bibinfo{pages}{027201} (\bibinfo{year}{2003})

\bibitem[{\citenamefont{ParkerChua}(1989)\citenamefont{Parker and Chua}}]{parkerchua}
\bibinfo{author}{\bibfnamefont{T.S.}~\bibnamefont{Parker}}
  \bibnamefont{and}
  \bibinfo{author}{\bibfnamefont{L.~O.} \bibnamefont{Chua}},
  \emph{\bibinfo{title}{Practical numerical algorithms for chaotic systems}}
(\bibinfo{publisher}{Springer-Verlag}, \bibinfo{address}{Berlin},
  \bibinfo{year}{1998}).

\bibitem[{\citenamefont{Eckmann and Ruelle}(1985)}]{eckmann}
\bibinfo{author}{\bibfnamefont{J.P.}~\bibnamefont{Eckmann}} \bibnamefont{and}
  \bibinfo{author}{\bibfnamefont{D.}~\bibnamefont{Ruelle}},
  \bibinfo{journal}{Rev.Mod.Phys.} \textbf{\bibinfo{volume}{57}},
  \bibinfo{pages}{617} (\bibinfo{year}{1985}).

\bibitem[{\citenamefont{Gao and Zheng}(1994)}]{gao}
\bibinfo{author}{\bibfnamefont{J.}~\bibnamefont{Gao}} \bibnamefont{and}
  \bibinfo{author}{\bibfnamefont{Z.}~\bibnamefont{Zheng}},
  \bibinfo{journal}{Phys. Rev. E} \textbf{\bibinfo{volume}{49}},
  \bibinfo{pages}{3807} (\bibinfo{year}{1994}).

\bibitem[{\citenamefont{R{\"u}melin}(1982)}]{rumelin}
\bibinfo{author}{\bibfnamefont{W.}~\bibnamefont{R{\"u}melin}},
  \bibinfo{journal}{SIAM J. Numer. Anal.} \textbf{\bibinfo{volume}{19}},
  \bibinfo{pages}{604} (\bibinfo{year}{1982}).

\bibitem[{\citenamefont{Hansen and Penland}(2006)}]{hansen}
\bibinfo{author}{\bibfnamefont{J.}~\bibnamefont{Hansen}} \bibnamefont{and}
  \bibinfo{author}{\bibfnamefont{C.}~\bibnamefont{Penland}},
  \bibinfo{journal}{Monthly Weather Rev.} \textbf{\bibinfo{volume}{134}},
  \bibinfo{pages}{3006} (\bibinfo{year}{2006}).

\bibitem[{\citenamefont{Kloeden and Platen}(1992)}]{kloeden}
\bibinfo{author}{\bibfnamefont{P.}~\bibnamefont{Kloeden}} \bibnamefont{and}
  \bibinfo{author}{\bibfnamefont{E.}~\bibnamefont{Platen}},
  \emph{\bibinfo{title}{Numerical Solutions of Stochastic Differential
  Equations}} (\bibinfo{publisher}{Springer-Verlag}, \bibinfo{address}{Berlin},
  \bibinfo{year}{1992}).

\bibitem[{\citenamefont{Hanggi}(1982)}]{hanggi}
\bibinfo{author}{\bibfnamefont{P.}~\bibnamefont{H{\"a}nggi}} \bibnamefont{and}
\bibinfo{author}{\bibfnamefont{H.}~\bibnamefont{Thomas}},
  \bibinfo{journal}{Phys. Rep.} \textbf{\bibinfo{volume}{88}},
  \bibinfo{pages}{207} (\bibinfo{year}{1982}).

\bibitem[{\citenamefont{Malik et~al.}(2010)\citenamefont{Malik, Ashok, and
Balakrishnan}}]{malik}
\bibinfo{author}{\bibfnamefont{N.}~\bibnamefont{Malik}},
  \bibinfo{author}{\bibfnamefont{B.}~\bibnamefont{Ashok}},
  \bibnamefont{and} \bibinfo{author}{\bibfnamefont{J.}~\bibnamefont{Balakrishnan}},
  \bibinfo{journal}{Eur. Phys. J. B} (\bibinfo{year}{2010}).
  \bibinfo{howpublished}{(DOI: 10.1140/epjb/e2010-00073-x)} 

\bibitem[{\citenamefont{Singer and Gray}(1995)}]{singer}
\bibinfo{author}{\bibfnamefont{W.}~\bibnamefont{Singer}} \bibnamefont{and}
  \bibinfo{author}{\bibfnamefont{C.}~\bibnamefont{Gray}},
  \bibinfo{journal}{Annu. Rev. Neurosci.} \textbf{\bibinfo{volume}{18}},
  \bibinfo{pages}{555} (\bibinfo{year}{1995}).

\bibitem[{\citenamefont{Kreiter and Singer}(1996)}]{kreiter}
\bibinfo{author}{\bibfnamefont{A.~K.} \bibnamefont{Kreiter}} \bibnamefont{and}
  \bibinfo{author}{\bibfnamefont{W.}~\bibnamefont{Singer}}, in
  \emph{\bibinfo{booktitle}{Brain Theory: Biological Basis and Computational
  Theory of Vision}}, edited by
  \bibinfo{editor}{\bibfnamefont{A.}~\bibnamefont{Aertsen}} \bibnamefont{and}
  \bibinfo{editor}{\bibfnamefont{V.}~\bibnamefont{Braitenberg}}
  (\bibinfo{publisher}{Elsevier}, \bibinfo{address}{Amsterdam},
  \bibinfo{year}{1996}).

\bibitem[{\citenamefont{Gerstner et~al.}(1997)\citenamefont{Gerstner, Kreiter,
  Markram, and Herz}}]{gerstner}
\bibinfo{author}{\bibfnamefont{W.}~\bibnamefont{Gerstner}},
  \bibinfo{author}{\bibfnamefont{A.~F.} \bibnamefont{Kreiter}},
  \bibinfo{author}{\bibfnamefont{H.}~\bibnamefont{Markram}}, \bibnamefont{and}
  \bibinfo{author}{\bibfnamefont{A.~V.~M.} \bibnamefont{Herz}},
  \bibinfo{journal}{Proc. Natl. Acad. Sci. USA} \textbf{\bibinfo{volume}{94}},
  \bibinfo{pages}{12740} (\bibinfo{year}{1997}).

\bibitem[{\citenamefont{Singer}(1999)}]{singer2}
\bibinfo{author}{\bibfnamefont{W.}~\bibnamefont{Singer}},
  \bibinfo{journal}{Neuron} \textbf{\bibinfo{volume}{24}}, \bibinfo{pages}{49}
  (\bibinfo{year}{1999}).

\bibitem[{\citenamefont{Ritz and Sejnowski}(1997)}]{ritz}
\bibinfo{author}{\bibfnamefont{R.}~\bibnamefont{Ritz}} \bibnamefont{and}
  \bibinfo{author}{\bibfnamefont{T.~J.} \bibnamefont{Sejnowski}},
  \bibinfo{journal}{Curr. Opin. Neurobiol.} \textbf{\bibinfo{volume}{7}},
  \bibinfo{pages}{536} (\bibinfo{year}{1997}).

\end{thebibliography}
\end{document}